\documentclass{emulateapj}
\usepackage{times}
\newcommand{\etal}{{et al}\/.}

\begin{document}
\slugcomment{Draft of \today}
\shorttitle{AGN feedback in NGC\,6764}
\shortauthors{J.H. Croston \etal}
\title{{\it Chandra} evidence for AGN feedback in the spiral galaxy NGC\,6764}
\author{J.H. Croston\altaffilmark{1}, M.J. Hardcastle\altaffilmark{1},
  P. Kharb\altaffilmark{2}, R.P. Kraft\altaffilmark{3}, A. Hota\altaffilmark{4}}
\altaffiltext{1}{School of Physics, Astronomy and Mathematics, University of
Hertfordshire, College Lane, Hatfield AL10 9AB, UK}
\altaffiltext{2}{Department of Physics, Purdue University, 525 Northwestern Avenue, West Lafayette, IN 47907, USA}
\altaffiltext{3}{Harvard-Smithsonian Center for Astrophysics, 60 Garden Street, Cambridge, MA~02138, USA}
\altaffiltext{4}{Institute of Astronomy and Astrophysics, Academia Sinica, P.O. Box 23-141, Taipei 10617, Taiwan, R.O.C.}

\begin{abstract}
 We report the {\it Chandra} detection of X-ray emission spatially
 coincident with the kpc-scale radio bubbles in the nearby ($D_{L}
 \sim 31$ Mpc) AGN-starburst galaxy NGC\,6764. The X-ray emission
 originates in hot gas ($kT \sim 0.75$ keV), which may either be
 contained within the radio bubbles, or in a shell of hot gas
 surrounding them. We consider three models for the origin of the hot
 gas: (1) a starburst-driven galactic wind, (2) shocked gas associated
 with the expanding radio bubbles, and (3) gas heated and entrained
 into the bubbles by jet/ISM interactions in the inner AGN outflow. We
 rule out a galactic wind based on significant differences from known
 galactic wind systems. The tight correspondence between the brightest
 X-ray emission and the radio emission in the inner outflow from the
 Seyfert nucleus, as well as a correlation between X-ray and radio
 spectral features suggestive of shocks and particle acceleration,
 lead us to favour the third model; however, we cannot firmly rule out
 a model in which the bubbles are driving large-scale shocks into the
 galaxy ISM. In either AGN-driven heating scenario, the total energy
 stored in the hot gas is high, $\sim 10^{56}$ ergs, comparable to the
 energetic impact of low-power radio galaxies such as Centaurus A, and
 will have a dramatic impact on the galaxy and its surroundings.

\end{abstract}

\maketitle

\section{Introduction}
\label{intro}

Galaxy feedback processes are now thought to be an important
ingredient in galaxy formation models (e.g. Croton et al. 2006; Bower
et al. 2006), potentially solving the long-standing problem of
explaining the deficit of galaxies at the low and high mass ends of
the galaxy luminosity function relative to model predictions.
Disentangling the feedback contributions from AGN outbursts and from
star formation to the energetics of gas in galaxies, galaxy groups and
clusters is a key problem in galaxy evolution, as it is clear that
both can inject large quantities of energy into their surroundings.
While there is growing evidence that AGN outbursts are the dominant
feedback process operating at the high-mass end of the galaxy
luminosity function, it might be expected that at lower masses, and
particularly in late-type galaxies, the energy input from star
formation processes will dominate.

Recent work has shown that kpc-scale radio bubbles connected to an
active nucleus can be found in many Seyfert galaxies (e.g. Gallimore
et al. 2006; Kharb et al. 2006). Although it has been argued that
these bubbles may be powered by stellar winds (e.g. Baum et al. 1993),
the fact that no convincing examples exist in galaxies without an AGN
strongly suggests that they are inflated by the active nucleus (e.g.
Hota \& Saikia 2006, hereafter HS06). Many of the known Seyfert radio
bubbles are predicted to be overpressured with respect to their
surroundings (e.g. Capetti et al. 1999), and therefore may be shocking
the surrounding medium, as has been observed for kpc-scale radio lobes
in massive elliptical galaxies (e.g. Centaurus A: Kraft et al. 2003;
NGC\,3801: Croston et al. 2007).

\begin{figure}[!t]
\begin{center}
\plotone{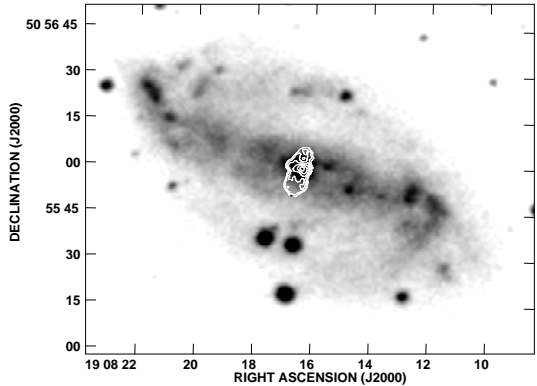}
\caption{1.4-GHz radio contours of NGC\,6764 overlaid on a DSS-2
optical image of the host galaxy. The radio map is made from VLA
archival data as published in Hota \& Saikia (2006). Contour levels
are $(1.5 \times 10^{-4}) \times 1, 2, 4,...,512$ Jy beam$^{-1}$.}
\label{ropt}
\end{center}
\end{figure}

In order to search for evidence for galaxy feedback associated with
kpc-scale radio bubbles in spiral galaxies, we carried out a {\it
Chandra} observation of the nearby Seyfert 2/LINER galaxy NGC\,6764 ,
which is a barred spiral galaxy (type SB(s)bc) with non-thermal radio
bubbles perpendicular to the plane of the galaxy disk that extend for $\sim
1$ kpc (Fig.~\ref{ropt}). Here we report on the results of this
observation.

Throughout this paper, we adopt a cosmology with $H_{0} = 70$ km
s$^{-1}$ Mpc$^{-1}$, $\Omega_{M} = 0.3$ and $\Omega_{\Lambda} = 0.7$.
We adopt a luminosity distance for NGC\,6764 of 31.3 Mpc, obtained by
correcting the heliocentric velocity of 2146 km s$^{-1}$ to the CMB
frame of reference. This gives an angular scale of $1^{\prime\prime} =
0.15$ kpc at the distance of NGC\,6764.

\section{Data analysis}

We observed NGC\,6764 with {\it Chandra} (ACIS-S) for 20 ks on 2008
January 20th.  The observation was taken in VFAINT mode to minimize
the background level. The data were reprocessed from the level 1
events file with {\sc ciao} 3.4 and CALDB 3.4.2, including VFAINT
cleaning. The latest gain files were applied and the 0.5-pixel
randomization removed using standard techniques detailed in the {\sc
ciao} on-line documentation\footnote{http://asc.harvard.edu/ciao/}. An
inspection of the lightcurve for our observation using the {\it
  analyze\_ltcrv} script showed that there were no periods of high
background level, and so no additional GTI filtering was applied. 

\begin{figure*}
\begin{center}
\caption{Left: 0.5 - 5.0 keV image of the {\it Chandra} data, smoothed
  with a Gaussian of 0.6$^{\prime\prime}$ FWHM, showing emission
  associated with the radio bubbles; right: X-ray emission in colour
  smoothed with a Gaussian of 1.8$^{\prime\prime}$ FWHM, with radio
  contours overlaid. Contour levels are as for Fig.~\ref{ropt}.}
\label{data}
\vskip 10pt
\plottwo{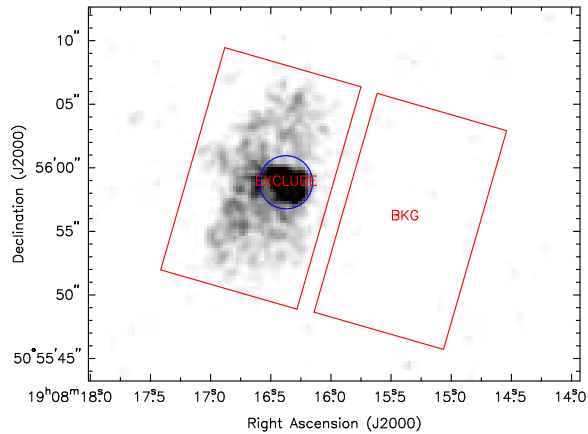}{f2b.eps}
\end{center}
\end{figure*}

We produced a 0.5 -- 5 keV filtered image to examine the X-ray
emission associated with the galaxy, presented in Fig.~\ref{data}. The
most prominent feature in the image is the region of extended
emission, which is located exactly coincident with the bubbles of
radio emission shown in the right hand panel. In addition, there is a
region of brighter X-ray emission in the centre of the emitting region
on scales of $\sim 4^{\prime\prime}$, coincident with the AGN and
extending to the west, while becoming broader. This bright region
coincides with a region of higher surface brightness radio emission,
as shown in the higher resolution 5-GHz radio map in Fig.~11 of HS06.

We extracted spectra from the Seyfert nucleus, the bright central
region, the two lobes (both separately and together to improve the
signal-to-noise ratio) and from a large region encompassing the galaxy
disk, using the {\it specextract} script, which also builds the
appropriate response files. Local background regions adjacent to or
surrounding the source extraction regions were used. Spectra were
grouped to 20 counts per bin after background subtraction prior to
spectral fitting, which was carried out using {\sc xspec}. We assumed
a fixed Galactic absorption of $N_H = 6.0 \times 10^{20}$ cm$^{-2}$ in
our spectral fitting (Dickey \& Lockman 1990), except where otherwise
specified.

\section{Results}

\subsection{Radio-related X-ray emission}

\begin{figure}
\begin{center}
\caption{ACIS-S counts spectrum for the entire bubble region in the
energy range 0.4 -- 7.0 keV, with best-fitting {\it mekal} model, as
described in the text, overplotted.}
\label{spec}
\plotone{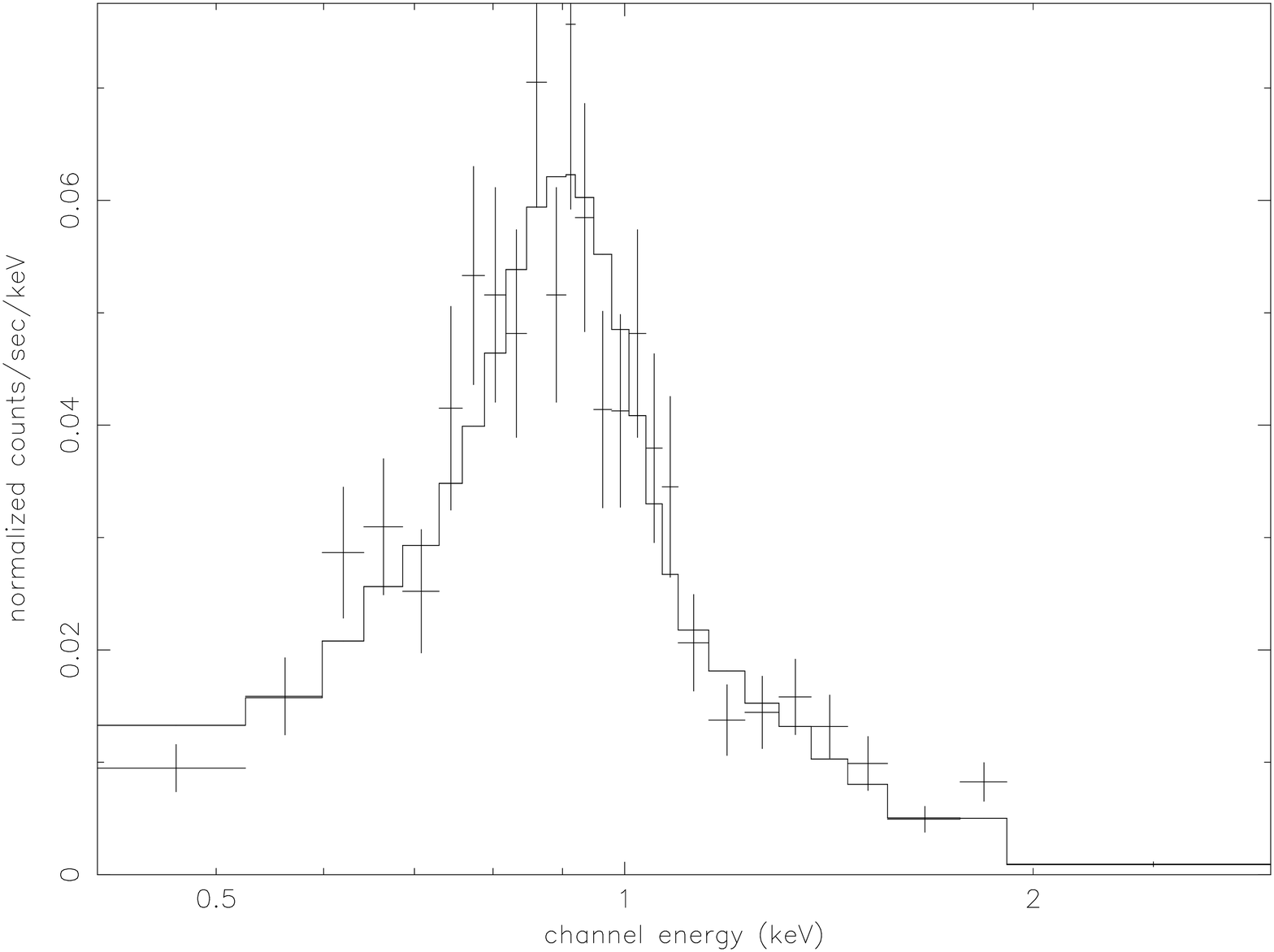}
\end{center}
\end{figure}

We measure a total of $\sim 1150$ net 0.5 -- 5 keV ACIS-S counts from
the regions coincident with the radio bubbles (excluding the bright
central region). In what follows we will refer to these X-ray emitting
regions as ``X-ray bubbles'' for simplicity but without any implied
geometry for the emitting fluid. We fitted a series of models to the
bubble spectra to distinguish between different origins for the X-ray
emission. We initially considered the spectrum for the entirety of the
bubble regions, using a local background region (source and background
extraction regions are shown in Fig.~\ref{data}). We fitted both
single power law and single {\it mekal} model (with Galactic
absorption, as described above, and free abundance for the thermal
model). The power-law model gave an unacceptable fitting statistic
($\chi^{2} = 181$ for 25 d.o.f.), and allowing the value of Galactic
column density to vary as a proxy for any absorption intrinsic to the
galaxy did not allow an acceptable fit to be obtained. We obtained a
good fit with the {\it mekal} model, with best-fitting values of $kT =
0.75\pm0.05$ keV and $Z = 0.13^{+0.05}_{-0.03}$ Z$_{\sun}$ and
$\chi^{2} = 25.5$ for 24 d.o.f. This model fit is shown in
Fig.~\ref{spec}. Allowing the total absorbing column to vary to
account for any intrinsic absorption does not significantly change the
best-fitting temperature or abundance values (although the total
absorbing column density was left free, it did not fall below the
assumed Galactic column). We note that the low abundance values may be
due to fitting a single-temperature model to a multi-temperature
plasma (e.g. Buote et al. 2000), and so are likely to be unreliable.
Given the very poor fitting statistics obtained for a single power-law
model, we can firmly rule out a non-thermal origin for the X-ray
emission associated with the radio bubbles. We added a power-law
component to the spectrum to investigate whether some contribution
from non-thermal emission is present; however, this did not
significantly affect the fit statistic or the parameters of the {\it
mekal} component. While we cannot rule out some contribution from
non-thermal emission, the X-ray emission predominantly originates from
gas of $kT \sim 0.75$ keV. We measure a total, unabsorbed 0.5 -- 2.0
keV flux from the X-ray bubbles of $1.2 \times 10^{-13}$ erg cm$^{-2}$
s$^{-1}$, corresponding to a luminosity of $\sim 1.4 \times 10^{40}$
erg s$^{-1}$.

To investigate whether there is any spatial variation in the
properties of the X-ray-emitting gas, we examined spectra of the
regions corresponding to the northern and southern bubbles separately.
The properties of the two bubbles appear to be fairly similar both to
each other and to the results for the joint spectrum discussed above.
In neither case is a power-law model an adequate fit (either with
fixed or free $N_{{\rm H}}$), and in both cases good fits could be
obtained with a single {\it mekal} model. For the northern bubble we
obtained a best-fitting temperature of $kT = 0.81^{+0.08}_{-0.07}$ keV
and $Z = 0.14^{+0.09}_{-0.05}$ Z$_{\sun}$ with $\chi^{2} = 10.3$ for 9
d.o.f., and for the southern bubble we obtained a best-fitting
temperature of $kT = 0.64^{+0.06}_{-0.05}$ keV and $Z =
0.12^{+0.06}_{-0.04}$ Z$_{\sun}$ with $\chi^{2} = 15.9$ for 13 d.o.f.
Therefore there is a small, but statistically significant difference
in the temperatures of the two bubbles, with the southern bubble
slightly cooler than the northern one. The origin of this small
difference in unclear, but could be related to differences in the
distribution of material being heated in the north and south. 

Finally, we also examined the spectrum of the brighter central region
of X-ray emission (Fig.~\ref{inner}; $2 - 4^{\prime\prime}$ scales) in
the same way. We found that for this emission neither a {\it mekal}
nor a power-law model gave an acceptable fit if Galactic absorption
was assumed; however, an acceptable fit could be obtained for a {\it
mekal} model if the value of $N_{\rm H}$ was permitted to vary (again
we varied the overall column density, but it did not drop below the
assumed Galactic value). We obtained a good fit with $N_{H} =
(2.0^{+1.5}_{-0.7}) \times 10^{21}$ cm$^{-2}$, $kT =
0.93^{+0.12}_{-0.18}$ keV, and abundance fixed at the value measured
for the outer regions, with $\chi^{2} = 16.7$ for 12 d.o.f. If the
abundance is fixed at solar abundance, we find a best fit with $N_{\rm
H} = (7.7^{+1.3}_{-1.0}) \times 10^{21}$ cm$^{-2}$ and $kT =
0.65^{+0.22}_{-0.11}$ keV, with $\chi^{2} = 18.3$ for 12 d.o.f. A
power-law model with free $N_{\rm H}$ did not give an acceptable fit,
and so we conclude that the brighter central X-ray emission also has a
thermal origin. The temperature of this gas is consistent with that in
the regions associated with the radio bubbles.

\subsection{The Seyfert nucleus}
\label{sey}
We extracted a spectrum from a 1.5$^{\prime\prime}$ radius circle
centred on the Seyfert nucleus, which we assumed to be located at the
peak of the X-ray emission. This position is in agreement with the
position of the Seyfert nucleus determined from optical observations
(Clements 1981) to within the accuracy of {\it Chandra}'s astrometry.
We found that single power-law models were unacceptable whether the
overall absorption was fixed at the Galactic value or left as a free
parameter as a proxy for a contribution from intrinsic absorption. A
single {\it mekal} model was also unacceptable with either Galactic or
free absorption. We therefore fitted a model consisting of a {\it
mekal} component plus an intrinsically absorbed power law. Since the
number of spectral bins is fairly low, the power-law index was fixed
at $\Gamma = 1.5$, a typical value for nuclear emission (a steeper
value, e.g. $\Gamma = 2.0$, does not significantly alter the results).
We obtained a good fit for this model, with $\chi^{2} = 10.1$ for 7
d.o.f., $kT = 0.92^{+0.16}_{-0.05}$ keV, and $N_{H} =
(2.9^{+3.6}_{-1.8}) \times 10^{21}$ cm$^{-2}$. For this best-fitting
model, we obtained an unabsorbed 2.0 -- 10.0 keV flux for the
power-law component of $(3.4^{+1.0}_{-1.2}) \times 10^{-14}$ ergs
cm$^{-2}$ s$^{-1}$, which corresponds to a 1-keV flux density of
$4.0^{+1.1}_{-1.4}$ nJy and implies a 2.0 - 10.0 keV unabsorbed
nuclear luminosity of $(4.0^{+1.2}_{-1.4}) \times 10^{39}$ ergs
s$^{-1}$. The uncertainities here take account of the large
uncertainty in the absorbing column. The unabsorbed 0.5 - 2.0 keV flux
of the thermal component is $(3.0^{+0.6}_{-0.5}) \times 10^{-14}$ ergs
cm$^{-2}$ s$^{-1}$, corresponding to a luminosity of
$(3.5^{+0.8}_{-0.6}) \times 10^{39}$ ergs s$^{-1}$. The total
unabsorbed nuclear 2.0 -- 10.0 keV luminosity is $\sim 4.4 \times
10^{39}$ ergs s$^{-1}$.

It is common for type 2 Seyfert nuclei to have strong intrinsic
absorption, artificially lowering the inferred X-ray luminosity [e.g.
Awaki et al. (2000), Terashima \& Wilson (2001), Matt et al. (2004),
Matsumoto et al. (2004)]. A more conservative upper limit can be
obtained by assuming a typical value of the intrinsic absorption for
Seyfert 2s where detailed X-ray analysis is available. We therefore
also fitted a model with a {\it mekal} plus heavily absorbed power
law, with $N_{H} = 3 \times 10^{23}$ cm$^{-2}$ [the value measured for
the Seyfert 2 M51 by Terashima \& Wilson (2001)] in order to determine
a conservative upper limit on the intrinsic nuclear luminosity. Fixing
the power-law normalisation at the upper limit allowed by the
best-fitting model, we found that the data are consistent with a
heavily obscured power-law component with a 2.0-10.0 keV luminosity as
high as $\sim 2 \times 10^{41}$ erg s$^{-1}$. Hence we cannot rule out
the presence of an intrinsically bright nucleus that is heavily
obscured.

\section{Energetics of the X-ray-emitting gas and radio bubbles}
\label{energetics}
As shown in Fig.~\ref{data}, the X-ray emission appears to trace the
1.4-GHz radio structure in NGC\,6764 very closely. There is evidence
for a surface-brightness deficit in both the X-ray and radio emission
in the centre of the northern bubble, and to a lesser extent in the
centre of the southern bubble; however, the X-ray structure is not
strongly edge-brightened as seen in shocked shells surrounding the
kpc-scale lobes of radio galaxies (e.g. Kraft et al. 2003; Croston et
al. 2007). We cannot distinguish between a model where the hot gas is
inside the radio bubbles and one where it is in shell structures
surrounding them based on the X-ray morphology alone.

To investigate the energetics of the bubbles we therefore considered
two possible scenarios: (1) the X-ray-emitting gas fills the radio
bubbles, (2) the X-ray-emitting gas is located in shells surrounding
the radio bubbles. Based on the outer extent of the X-ray emission
from the southern bubble (which has a more spherical appearance), we
assumed an outer radius for the emitting region of
$4.2^{\prime\prime}$, which corresponds to 0.7 kpc. For scenario (1)
we assumed a sphere of this radius, and for scenario (2) we assumed a
spherical shell with thickness 0.1 kpc. Assuming a constant density
and a filling factor of unity in the emitting region, these geometries
lead to inferred electron densities of 0.19 cm$^{-3}$ and 0.30
cm$^{-3}$, for scenarios 1 and 2, respectively. For the measured
temperature of the southern bubble X-ray emission, these correspond to
pressures of $(4 - 7) \times 10^{-12}$ dyne cm$^{-2}$. For the
northern bubble, we assumed an outer radius of $4.0^{\prime\prime}$,
and the same shell thickness for scenario 2, which led to inferred
electron densities of 0.17 cm$^{-3}$ and 0.27 cm$^{-3}$ for scenarios
1 and 2, respectively, and to pressures consistent with those of the
southern bubble. For lower filling factors, the gas densities (and
pressures as inferred below) will increase. The total gas mass is
$\sim 10^{7}$ M$_{\sun}$, a small fraction of the ISM mass likely to
be contained in the galaxy's bulge.

\begin{figure}
\caption{Surface brightness profiles for the northern (l) and southern
  (r) bubbles obtained in a slice towards the bubble edge. Centre of
  the profile is the bubble centre in each case (not the galaxy
  nucleus).}
\label{profs}
\plottwo{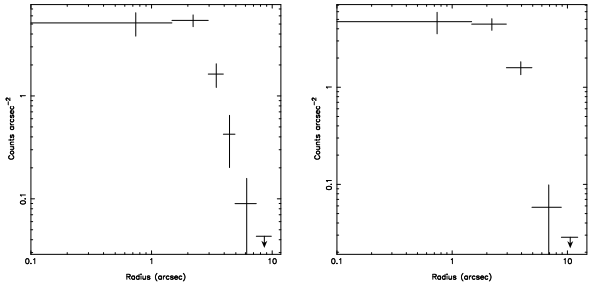}{f4b.eps}
\end{figure}
\begin{figure*}
\caption{Left: Close-up of the X-ray emission from the inner
  jet/outflow region from the 0.5 - 5.0 keV image smoothed with a
  Gaussian of 0.6$^{\prime\prime}$ FWHM, with 5-GHz contours overlaid
  from a map made from archival data as published in Hota \& Saikia
  (2006). Contour levels are $(8.0 \times 10^{-5}) \times 1, 2, 4,
  ..., 512$ Jy beam$^{-1}$; right: hardness ratio map, as described in
  the text. White is hard and red is soft. There is a statistically
  significant increase in hardness to the western edge associated with
  a region of flatter spectrum radio emission.}
\label{inner}
\vskip 10pt
\plotone{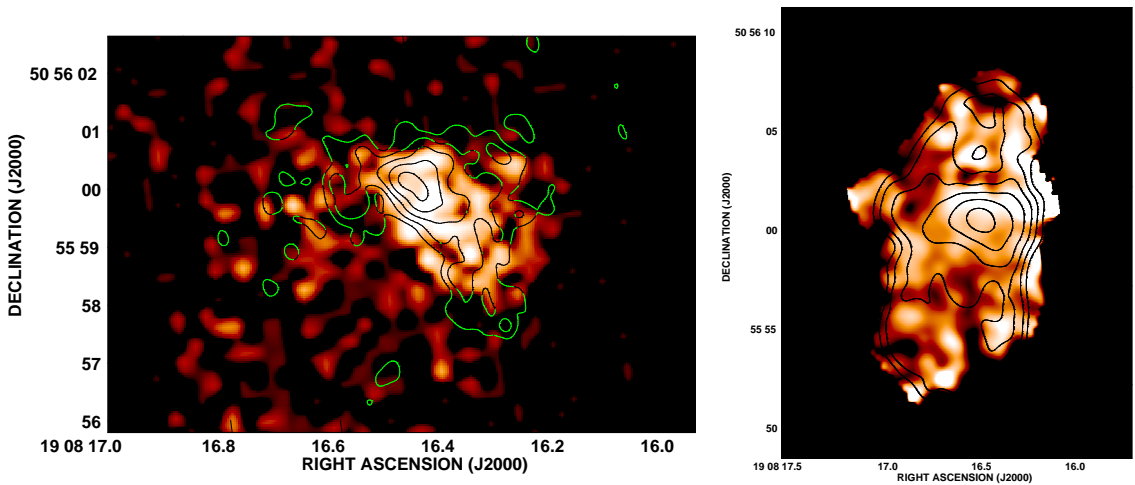}
\end{figure*}

We determined the equipartition internal pressures of the two bubbles
using measurements of the 1.4-GHz flux density for each lobe to
normalize the synchrotron spectrum. We assumed a filling factor of
unity, a broken power-law electron distribution with initial electron
energy index, $\delta$, of 2.1, $\gamma_{{\rm min}} = 10$ and
$\gamma_{{\rm max}} = 10^{5}$, and a break at $\gamma_{{\rm break}} =
10^{3}$, and modelled the lobes as spheres of radius 4.0 and
4.2$^{\prime\prime}$, respectively, for the northern and southern
bubbles. The resulting equipartition internal pressures were $\sim 1.2
\times 10^{-13}$ dyne cm$^{-2}$ for both bubbles, which are a factor
of $\sim 30 - 50$ times lower than the pressure in the hot gas. Hence
if the hot gas is external to the radio bubbles, an additional source
of pressure in the bubbles is required to balance that in the hot gas.
However, it is known that external pressures from hot gas surrounding
most low-power radio-galaxy lobes are significantly higher than
internal equipartition pressures (e.g. Morganti et al. 1988; Croston
et al. 2003, 2008), and so this is not on its own a strong argument
against the gas being outside the radio bubbles. In particular, the
external pressures of the shocked shells in NGC\,3801 and Centaurus A
(Croston et al. 2007; Kraft et al. 2003) are factors of $\sim 70$ and
$\sim 15$ times higher, respectively, than the internal equipartition
radio-lobe pressures.

Based on the two scenarios for the geometry of the X-ray-emitting
material discussed above, we determined the total thermal energy
stored in the hot gas to be $(3 - 5) \times 10^{55}$ ergs, which is
comparable to the energy stored in the shocked shells surrounding the
lobes of the radio galaxy NGC\,3801 ($\sim 8 \times 10^{55}$ ergs).

We also determined the physical properties of the central region of
brighter, thermal X-ray emission. As shown in Fig.~\ref{inner}, this
region corresponds closely to a brighter region of radio emission. The
narrowest part of the brighter region coincides with the position of
the Seyfert nucleus, leading HS06 to interpret this emission as a
radio jet oriented towards the western edge of the large-scale radio
bubbles. The X-ray and radio emission from this brighter region
corresponds very closely, and it has the appearance of a poorly
collimated outflow. If this is indeed an outflow, then it is likely to
be supplying material to the large-scale bubbles. We therefore also
investigated the energetics of this inner region. Assuming a spherical
region of radius 1.85$^{\prime\prime}$, we find an electron density
between 0.76 -- 0.80 cm$^{-3}$ corresponding to a gas pressure of
$(8.3 - 11) \times 10^{-12}$ dyne cm$^{-2}$ (the ranges given here
take into account a range in metallicity from [0.15 -- 1]Z$_{\sun}$).
Hence the inner gas is overpressured by as much as a factor two
relative to the outer X-ray bubbles, and should be expanding and
possibly heating the outer gas through weak shocks. It is also
possible that the radio bubble is providing pressure support for this
higher pressure gas, or that the gas is being compressed by the
expanding radio bubble. We discuss the implications of this expanding
hot gas in the centre of the galaxy, and its relation to the larger
X-ray bubbles in the sections that follow.

\section{Origin of the X-ray emission: galactic wind or jet-driven shocks?}

The two most likely explanations for the origin of the X-ray emission
in NGC\,6764 are: (I) hot gas in a galactic wind, similar to the X-ray
emission seen in typical nearby starburst galaxies, or (II) emission
from material that has been heated by the expanding radio plasma.
However, there are some clear problems with both scenarios, as neither
can fully explain the very tight correspondence between radio and
X-ray morphology. A third possibility, suggested by this
correspondence, is that both the radio and X-ray emission are produced
by the same particles. However, the steep radio spectral index of the
bubbles is not consistent with a thermal origin for the radio
emission, and the X-ray spectra of the bubbles are clearly
inconsistent with a non-thermal origin. The observed flux level is
also much higher than would be predicted by an inverse-Compton model,
so that this model can be firmly ruled out. Therefore, we can find no
obvious explanation in which the X-ray emission is produced by the
same particles as the radio emission. Below we consider possibilities
(I) and (II) in more detail. We consider two scenarios for Model II: a
model in which the expanding bubbles are driving large-scale shocks
into the ISM (Model IIa), and one in which jet/ISM interactions in the
inner parts of the outflow are heating the ISM and causing it to be
entrained into the bubbles (Model IIb).

\subsection{Model I: hot gas from a galactic wind}

As NGC\,6764 is a galaxy with known starburst activity, the presence
of a galactic wind (e.g. Veilleux et al. 2005) producing the observed
X-ray emission must be considered. The morphology of the X-ray
emission from NGC\,6764 is similar to some of the known examples of
X-ray-detected galactic winds in nearby galaxies (e.g. Strickland et
al. 2004b). For example, the X-ray structures in the starburst
galaxies M82 and NGC\,1482 have similar opening angles to the bubbles
in NGC\,6764. In both of these cases the X-ray-emitting gas is
coincident with H$\alpha$ filaments extending out of the galaxy disk,
which is also the case for NGC\,6764 (see Fig. 12 of HS06). However,
radio bubbles coincident with the X-ray-emitting gas are not present
in either M82 or NGC\,1482. We note also that the X-ray morphology of
these two systems is not consistent with a uniformly filled sphere or
truncated cone, as it is slightly brighter towards the edges rather
than in the centre.

NGC\,6764 shows strong evidence for circumnuclear starburst activity
(e.g. Eckart et al. 1991, 1996; Schinnerer et al. 2000). There is also
evidence that the nuclear starburst is producing an outflow: e.g. Leon
et al. (2007) estimated, based on CO observations, that $\sim 4 \times
10^{6}$ M$_{\sun}$ of molecular gas is outflowing at speeds of $\sim
25$ km s$^{-1}$. However, Eckart et al. (1991) estimated a total star formation
rate of $\sim 4$ M$_{\sun}$ yr$^{-1}$, with a supernova rate of $\sim
1.2 \times 10^{-2}$ yr$^{-1}$, which is significantly lower than the
SFR of $13 - 33$ M$_{\sun}$ yr$^{-1}$ for M82 estimated by F\"{o}rster
Schreiber et al. (2003). The kinetic energies of the outflowing
material detected in HI absorption (HS06) and CO emission (Leon et al.
2007) are $8.5 \times 10^{52}$ ergs and $2.4 \times 10^{53}$ ergs,
respectively, which are both several orders of magnitude below the
thermal energy of the X-ray emitting gas. Leon et al. estimated the
typical energies available from supernovae and stellar winds given its
inferred star formation rate, finding values in the range $10^{54} -
10^{55}$ ergs. It is therefore unclear that sufficient energy is
available from starburst activity to power the observed X-ray
emission.

The median 0.3 -- 2.0 keV X-ray luminosity for galactic winds for the
starburst galaxies in the sample of Strickland et al. (2004a, 2004b)
is $2.5 \times 10^{39}$ erg s$^{-1}$, which is an order of magnitude
lower than the X-ray luminosity of the emission from NGC\,6764 in the
same energy range. The highest luminosity wind in their sample has a
0.3 -- 2.0 keV X-ray luminosity of $L_{X} = 8.3 \times 10^{39}$ erg
s$^{-1}$, a factor of 2 lower than that of NGC\,6764. While NGC\,6764
is therefore more luminous than other known galactic winds, its
luminosity is not so extreme as to preclude a similar origin. However,
the X-ray emission from the bubbles of NGC\,6764 (excluding the bright
central region) originates from a smaller volume (e.g. the height in
kpc above the disk to which the emission extends is a factor of $\sim
3$ times smaller in NGC\,6764 than in M82), hence the volume
emissivity is considerably higher.

The diffuse $H\alpha$ emission from NGC\,6764 has a total luminosity
of $(6 - 8) \times 10^{40}$ erg s$^{-1}$ (Zurita et al. 2000), giving
a value of $L_{X}/L_{H\alpha} \sim 3 - 4$. In contrast, the ratio of
$L_{X}/L_{H\alpha}$ for the starburst wind in M82 is $\sim 19$ (using
the $H\alpha$ luminosity measured by McCarthy et al. (1987). We note
also that the ``mean'' temperatures for the galactic winds in the
Strickland et al. sample are typically significantly lower than that
of the gas in NGC\,6764, although these result from two-temperature
fits where the hotter component has temperatures similar to our
measured value for NGC\,6764. We find no evidence for a cooler
component of hot gas in NGC\,6764 (in two-temperature fits, the second
component tends to higher temperatures of $\sim 1$ keV with no
significant improvement in the fitting statistic).

Another important difference between the X-ray emission in NGC\,6764
and galactic winds such as M82 is the sharp decrease in surface
brightness at its northern and southern boundaries. Rasmussen et al.
(2004) compare the inferred density slopes at the outer boundaries of
three starburst winds, M82, NGC\,253 and the dwarf galaxy NGC,1800,
finding slopes of $\alpha \sim 1$, where $n(r) \propto r^{-\alpha}$.
In constrast, NGC\,6764 shows an initially flat surface brightness
profile (see Fig.~\ref{profs}), dropping steeply at distances between
2 and 5$^{\prime\prime}$ for both the northern and southern bubbles,
with a slope that corresponds to $\alpha \sim 2.5$. This is a steeper
decrease than would be expected for a freely expanding wind (e.g.
Chevalier \& Clegg 1985). 

The presence of hot, overpressured gas in the inner regions of the
galaxy is consistent with the presence of a galactic wind; however,
its one-sided structure and East-West orientation may be difficult to
reconcile with a galactic wind model. We are not aware of any other
galactic wind systems in which a central hot-gas outflow with this
type of structure has been observed. An alternative is a photoionized
wind that is driven directly by the AGN. Such a model cannot be ruled
out on energetic grounds alone: our upper limit on the possible hidden
nuclear X-ray luminosity is higher than the energy required to power
the bubbles for realistic timescales of bubble inflation (e.g. $10^{6}
- 10^{7}$ years). However, as for the starburst-driven wind model, the
AGN wind model does not appear consistent with the steep decrease in
surface brightness of the X-ray bubbles in NGC\,6764. The origin of
the radio emission is also unclear in this model, and the E-W
orientation of the brighter central outflow may be problematic for an
AGN wind explanation.

We conclude that there are important differences between the
properties of the X-ray emission and starburst activity of NGC\,6764
compared to those of well-studied examples of starburst-driven galactic
winds. Most crucially, the close correspondence between extended
non-thermal radio emission and X-ray emission is not seen in any of
the starburst wind systems. HS06 argue that the lack of such radio
bubbles in any galaxies that do not possess an AGN indicates that the
bubbles are likely to be powered by the AGN radio jet, although their
dynamics and evolution may be affected by the presence of a galactic
wind.

\subsection{Model IIa: shock heating by the expanding radio bubbles}

The properties of the hot gas associated with the radio bubbles in
NGC\,6764 are similar to those of the shocked gas associated with the
kpc-scale radio lobes of Centaurus A and NGC\,3801. The temperatures
we measure are similar to those seen in NGC\,3801, and the gas
densities and pressures are comparable to those measured in the
shocked shells of gas surrounding Centaurus A [assuming the thermal
interpretation of Kraft et al. (2003)]. However, there is no strong
evidence for edge-brightening of the X-ray emission -- in fact both
the radio and X-ray emission appear to have similar structures, with a
slight surface brightness deficit towards the centre of the bubbles,
particularly in the north. It would be possible to contrive a shock
geometry in which there was more shocked material in front of or
behind the bubbles relative to that at the sides, thus counteracting
the effect of edge-brightening; however, the smoothness of the X-ray
emission argues against a model in which the shocked material is very
inhomogeneous.

We first considered the case where the X-ray emission is produced from
shocked shells of gas surrounding the radio bubbles. Assuming a strong
adiabatic shock, the expected density contrast between shocked and
unshocked gas should be a factor of 4. Therefore a shocked gas model
requires a surrounding medium with $n_{e} \sim 0.08$ cm$^{-3}$.  As X-ray surface brightness depends
primarily on the gas density, we can use an upper limit on the X-ray
surface brightness from regions surrounding the X-ray emission
associated with the radio bubbles to investigate whether a hot gas
medium could be being shocked. We considered a rectangular region
covering the eastern half of the southern radio bubble and a matched
background region next to the X-ray bright region. We found a surface
brightness of $\sim 9.4\pm0.1$ net ACIS-S 0.5 -- 5 keV counts
arcsec$^{-2}$ in the source region. If hot gas with a density a factor
of 4 lower than that in the bright regions were present, we would
expect it to have a surface brightnes a factor $\sim 16$ times lower,
e.g. $S_{X} \sim 0.59\pm0.01$ counts arcsec$^{-2}$ (if the unshocked
material is cooler this prediction decreases somewhat, e.g. to 0.5
counts arcsec$^{-2}$ for $kT = 300$ eV). We measure an upper limit of
$\sim 0.1$ counts arcsec$^{-2}$ from the region external to the
southern radio bubble, well below this prediction. It therefore seems
unlikely that hot ionized plasma can be the medium being shocked.

The densities of molecular gas in the centre of NGC\,6764 are much
higher than our prediction for the density of the material being
shocked (e.g. $n_{e} \sim 3 \times 10^{3} - 3 \times 10^{4}$
cm$^{-3}$; Eckart et al. 1991). A possible candidate for material
being shocked is the warm, diffuse interstellar medium, which has
typical electron densities of $\sim 0.2$ cm$^{-3}$ (e.g. Walterbos
1998). If the bubbles extend perpendicular to the plane of the galaxy
disk, then they are likely to be probing regions of lower than average
ISM density. In addition, a weaker shock could result in a density
contrast lower than 4, and so this candidate for shocked material
cannot be ruled out on grounds of density. As seen in Fig.~12 of HS06,
the radio bubbles of NGC\,6764 are coincident with an extended region
of H$\alpha$-emitting gas (Zurita et al. 2000), with brighter
filaments associated with the radio bubbles, indicating that the radio
bubbles are likely to be embedded in diffuse gas.

The most likely large-scale shock scenario is one in which a strong
shock is propagating into a multiphase medium. We can use the
postshock temperature to estimate the shock speed, assuming that the
pre-shocked gas pressure is negligible. The inferred expansion speed
is $\sim 740$ km s$^{-1}$, which is comparable to the radio-lobe
expansion speeds measured from the shock properties in NGC\,3801
($\sim 850$ km s$^{-1}$), and would imply a total kinetic energy in
the shocked material of $\sim 3 \times 10^{55}$ ergs, similar to, but
slightly lower than the thermal energy stored in the shells. If the
shells are expanding at such a high speed, then the age of the X-ray
features must be low ($\sim 10^{6}$ years, assuming a sound speed
appropriate for a hot gas ISM), implying that either the features are
short-lived or the gas supply is being replenished. If the latter is
the case, then an outflow of gas from the galaxy is likely to be
required to supply the ionized gas halo, even if it does not supply
the energy to power the X-ray emission.

\subsection{Model IIb: Heating by jet/ISM interactions in the central
  outflow region}

The X-ray bright, overpressured region in the central few arcsec of
the galaxy suggests a second model for jet-related heating of the
X-ray-emitting gas. If this region is an outflow that is supplying
material for the large-scale bubbles, then it is clear from the good
radio and X-ray correspondence that it contains both hot gas and a
population of radio-emitting, non-thermal particles, presumably
supplied by a radio jet. The central X-ray/radio morphology is hard
to explain in Model IIa, although the X-ray emission could be caused
by expansion of the bubbles near to the nucleus. The most plausible
explanation for the close correspondence, assuming that the radio
plasma traces a nuclear outflow, is that the ISM is being heated and
compressed as it interacts with the radio-emitting outflow. The
interactions may also be responsible for decollimating the flow. If
the radio plasma and heated gas components can become well mixed in
the inner outflow, then it seems likely that they will remain mixed on
larger scales. Hence an alternative jet-heating scenario is one in
which the inner jet is disrupted by interactions with the clumpy ISM,
which also heat ISM gas and cause it to be mixed in with the jet
fluid. This is similar to the ``frustrated jet'' model discussed by
Gallimore et al. (2006), which predicts soft X-ray emission associated
with shocks and entrained material. While the energy is clearly
available from the jet to enable such heating on small scales to
occur, it is unclear how the details of such a heating and mixing
process can work in practice.

In this scenario the bubbles are filled with hot, thermal gas. In this
case we would expect significant depolarization of the radio
synchrotron emission. Using the radio data discussed above, we found
that the integrated fractional polarization of the bubbles is $<2 \%$
and $<5 \%$ at 1.4-GHz and 5-GHz, respectively. This is consistent
with the level of expected internal Faraday depolarization given the
gas density and equipartition magnetic field strength in the bubbles
(see Cioffi \& Jones 1980). While the level of polarization is
consistent with Model IIb, it is also consisent with an external
Faraday screen having the density and path length expected for shocked
gas surrounding the bubbles in Model IIa (for reasonable values of the
magnetic field strength and plausible assumptions about the cell size
in the Faraday screen) using the analysis of Burn (1966), and so we
cannot rule out Model IIa on this basis.

The right-hand panel of Fig.~\ref{inner} shows a hardness ratio map
obtained by dividing Gaussian smoothed images in the 1.0 -- 5.0 keV
and 0.4 -- 1.0 keV energy ranges. The region of harder X-ray emission
along the western edge of the bubbles is statistically significant (as
verified by comparing count ratios with uncertainties for the western
and eastern edges of the bubbles), and corresponds to a region of
flatter radio spectral index as shown in HS06. Gallimore et al. (1996)
propose a similar explanation for component C of the parsec-scale
jet in NGC\,1068 parsec-scale jet, which has a similar spectral index
to the regions we consider here and is associated with H$_{2}$0 maser
emission. These results suggest that the East-West outflow may be
shock heating the ISM in this region, leading to an increase in X-ray
temperature, and a flattening of the radio spectrum due to localized
particle acceleration. The radio and X-ray spectral structure
therefore offer support for a model in which shock heating plays at
least some role in heating the X-ray emitting gas. The edge-brightened
radio and X-ray structure of the bubbles suggests that shock heating
and particle acceleration may also be occurring at the edges of the
bubbles (although there is no flattening of the radio spectrum, or
hardening of the X-ray spectrum around the eastern edges). Magnetic
field compression offers an alternative explanation for the radio
structure, but does not explain the close correspondence with the
thermal X-ray emission. The surface brightness profiles of the bubbles
(Fig.~\ref{profs}) are too steep to be consistent with a uniformly
filled sphere of gas, supporting a model in which some compression and
possibly shocks are occurring at the edges of the bubbles.

Sutherland \& Bicknell (2007) presented simulations of jet/ISM
interactions in a non-homogeneous medium that show some similarities
with our observations: the radio and X-ray morphologies of the
large-scale bubbles in NGC\,6764 are somewhat similar to their
``energy-driven bubble'' phase, with the inner region having a
morphology that could be consistent with their initial ``flood and
channel'' phase, which they find persists in the inner regions of the
outflow during the energy-driven phase. In their simulations X-ray
emission from shocks driven out by the bubbles is also significant.
However, the simulations were for much more powerful jets and for
initial environmental conditions quite different to those of NGC\,6764
(including a static clumpy ISM), Nevertheless they may provide some
qualitative support for a jet/ISM interaction interpretation in
NGC\,6764.

If it is the radio jet and its environmental impact that are the
primary explanation for the bubbles, then it is perhaps surprising
that the inner jet/outflow is oriented in the East-West direction, not
North-South. The can be explained if the jet direction can change on
short timescales, perhaps due to the interactions with dense
surrounding material in the inner regions. Kharb et al. (2006)
proposed such a model for the Seyfert 2 galaxy Markarian 6, which
possesses an inner jet nearly perpendicular to its large-scale radio
bubbles, concluding that the jet direction changes on timescales of
$10^{5} - 10^{6}$ years. The radio and X-ray outflow properties of
NGC\,6764 are consistent with the picture described by Gallimore et
al. (2006) in which jet-ISM interactions in the inner kiloparsec lead
to frustration of the AGN jet. We conclude that this model is somewhat
more plausible than a large-scale radio-bubble driven shock (Model
IIa); however, neither model can be ruled out.

\subsection{Comparisons with other Seyfert radio-bubble systems}

HS06 presented a list of Seyfert radio-bubble systems similar to
NGC\,6764. {\it Chandra} X-ray studies of three of these systems are
published in the literature, NGC\,1068, M51 and NGC\,3079.
Below we discuss the similarities and differences between NGC\,6764
and each of these previously studied radio-bubble systems.

The Seyfert 2 galaxy NGC\,1068 has two asymmetric radio bubbles with a
total extent of $\sim 1$ kpc, which are connected to a radio-bright
nucleus. The {\it Chandra} study of Young et al. (2001) showed the
presence of bright extended X-ray emission associated with the
north-eastern radio bubble, with a lack of corresponding emission
associated with the south-western bubble (likely to due to high
absorption from the disk of the galaxy). The X-ray and radio
morphology of this system are consistent with similar behaviour to
NGC\,6764. Young et al. required more complex spectral models to
account for the X-ray emission associated with the north-east radio
bubble region; however, they were able to obtain spectra of much
higher signal-to-noise than ours for NGC\,6764, and so it is possible
that a single-temperature {\it mekal} model would also prove
inadequate if we were able to obtain similar quality data for
NGC\,6764. Young et al. suggest that compression of the ISM by the
radio ejecta is responsible for the bubble-related X-ray emission.

Terashima \& Wilson (2001) presented a {\it Chandra} study of the
extended emission from the Seyfert 2 galaxy M51, which possesses a
highly asymmetry two-sided radio bubble structure. As for NGC\,6764,
they found extended X-ray emission corresponding closely to the radio
structure. The X-ray emission is spectrally similar to that seen in
NGC\,6764 ($kT \sim 0.5 - 0.6$ keV), and interestingly a region of
harder X-ray emission is seen at the position where the radio jet is
seen to teminate, similar to what is seen in NGC\,6764. Terashima \&
Wilson adopt a shock heating model for M51, finding bubble expansion
velocities very similar to those inferred above for NGC\,6764 ($\sim
700$ km s$^{-1}$). 

NGC\,3079 is another composite starburst/AGN system, which possesses
two kpc-scale scale radio bubble structures . Although the radio
structures have been modelled as originating in a nuclear wind (e.g.
Duric \& Seaquist 1988), long-term VLBI monitoring shows emitting
regions whose structure and time variability is consistent with the
expectation for a jet interacting with a dense and clumpy surrounding
medium (Middelberg et al. 2007). Middelberg et al. also argue that
these interactions can explain the large-scale morphology of the radio
bubbles. Cecil et al. (2002) present {\it Chandra} observations of
NGC\,3079, which show filamentary X-ray structure coinciding with
H$\alpha$ filaments; however, there is also a region of X-ray emission
associated with the southern radio bubble, which is too faint for
spectral analysis. The optical and X-ray filaments to the north do not
coincide with the radio bubble, but Cecil et al. claim that there is
some evidence for shock heating of gas surrounding the VLBI jet. It is
clear that starburst winds are responsible for some of the X-ray
structure in NGC\,3079, but it remains possible that X-ray emission
heated by the expanding radio plasma is also present. The relationship
between the radio structure and the optical and X-ray filaments in
NGC\,3079 is unclear. 

Based on the comparisons above, it seems likely that the radio/X-ray
bubbles observed in NGC\,6764, NGC\,1068 and M51 are examples of the
same phenomenon, whereas the situation in NGC\,3079 may be more
complex. We conclude that the behaviour seen in NGC\,6764 may be
widespread in the Seyfert population, as such outbursts are likely to
be short-lived. We would expect similar X-ray properties for radio
bubbles in Seyfert 1 galaxies; however, due to the small size scales
of the lobes and the expected projection, bubble-related X-ray
emission is likely to be difficult to disentangle from the brighter
nuclear X-ray emission in these systems.

\section{Conclusions}

We have identified X-ray counterparts to the kpc-scale radio bubbles
in the Seyfert galaxy NGC\,6764, whose morphology traces very closely
the radio structures. The energy input from these hot gas features is
$\sim 10^{56}$ ergs, comparable to the energy input from shock heating
in the low-power radio galaxy NGC\,3801. The inferred rate of energy
input in NGC\,6764 is $\sim 10^{42}$ erg s$^{-1}$, and is expected to
have a significant impact on the galaxy and its surroundings. Based on
the presence of an inner outflow that appears to emanate from the
Seyfert nucleus, seen in both radio and X-ray emission, similarities
to X-ray features in previously studied AGN-jet systems, key
differences compared to starburst outflows, and the spatial
coincidence of a region of flatter radio spectrum and hard X-ray
spectrum, suggesting particle acceleration, we conclude that the
thermal emission is produced by heating of the ISM due to compression
and shocks caused by the AGN outflow. The morphology of the radio
outflow in NGC\,6764 leads us to favour a model in which jet/ISM
interactions in the inner part of the galaxy lead to heating of the
gas and its entrainment into the bubbles, as well as jet disruption;
however, we cannot rule out a model in which the bubbles are driving
large-scale shocks into the ISM. The existence of a population of
kpc-scale radio lobe sources in Seyfert galaxies (expected to be
short-lived) suggests that AGN-jet or AGN-jet+starburst powered galaxy
feedback as seen in NGC\,6764 could be important for the galaxy
population as a whole.

\acknowledgments

We thank the referee for helpful comments. We gratefully acknowledge
support from the Royal Society (research fellowship for MJH). This
work was partially supported by NASA grant G08-9108X.

\end{document}